\documentclass[12pt]{article} 
\pdfoutput=1
\usepackage{graphicx,psfrag,axodraw}
\usepackage{amsmath,amsfonts,amssymb,stmaryrd,latexsym}
\usepackage{url,multirow}
\newcommand{\be}{\begin{equation}}
\newcommand{\ee}{\end{equation}}
\newcommand{\bea}{\begin{eqnarray}}
\newcommand{\eea}{\end{eqnarray}}
\newcommand{\bear}{\begin{eqnarray}}
\newcommand{\eear}{\end{eqnarray}}

\newcommand{\ba}{\begin{array}}
\newcommand{\ea}{\end{array}}

\def\gev{\,{\rm GeV}}
\def\SU2{\rm SU(2)}
\def\U1{\rm U(1)}

\newcommand{\met}{ E_T\!\!\! \!\! \! \slash \;\, } 

\begin{document}

\baselineskip=18pt \pagestyle{plain} \setcounter{page}{1}

\vspace*{-1cm}

\noindent \makebox[13cm][l]{\small \hspace*{-.2cm}
FERMILAB-Pub-15-286-T
}{\small  PITT-PACC-1510}  \\  [-3mm]

\begin{center}

{\Large \bf  Heavy Higgs bosons and the 2 TeV $W'$ boson   } \\ [9mm]

{\normalsize \bf Bogdan A. Dobrescu$^\star$ and Zhen Liu$^{\star \diamond}$\\ [3mm]
{\small {\it
$\star$ Theoretical Physics Department, Fermilab, Batavia, IL 60510, USA   \\ [1mm]
$\diamond$  
Department of Physics and Astronomy, University of Pittsburgh, PA 15260, USA   }}\\ [2mm]
}

July 7, 2015; revised Aug 28, 2015
\end{center}


\begin{abstract}
The hints from the LHC for the existence of a $W'$ boson of mass around 1.9 TeV 
point towards a certain  $SU(2)_L\times SU(2)_R\times U(1)_{B-L}$ 
gauge theory with an extended Higgs sector. We show that the decays of the $W'$ boson into heavy Higgs bosons have sizable branching fractions.
Interpreting the ATLAS excess events in the search for same-sign lepton pairs plus $b$ jets as arising from 
$W'$ cascade decays, we estimate that the masses of the heavy Higgs bosons are in the 400--700 GeV range. 
\end{abstract}

\tableofcontents

\section{Introduction} \setcounter{equation}{0}

Using LHC data at $\sqrt{s} = 8$ TeV, the ATLAS and CMS Collaborations have reported deviations from the Standard Model (SM) 
of statistical significance between 2$\sigma$ and 3$\sigma$ in several final states, indicating mass peaks in the 1.8--2 TeV range   
\cite{Aad:2015owa}-\cite{Khachatryan:2014gha}.
The cross sections required for producing these mass peaks are consistent with the properties of a $W'$ boson in an
$SU(2)_L \times SU(2)_R \times U(1)_{B-L}$ gauge theory with right-handed neutrinos that have Dirac masses at the TeV scale \cite{Dobrescu:2015qna}.

The spontaneous breaking of $SU(2)\times SU(2)\times U(1)$  gauge groups requires an extended Higgs sector. For large regions of parameter space, 
the $W'$ boson has large branching fractions into heavy scalars from the Higgs sector \cite{Dobrescu:2013gza}\cite{Jinaru:2013eya}.
We show here that the $W'$ boson hinted by the LHC data may also decay into $H^+A^0$ and $H^+H^0$, where $H^+$, $A^0$ and $H^0$ are 
heavy spin-0 particles present in Two-Higgs-Doublet models. We compute the branching fractions for these decays and present some evidence that 
signals for the $W'\to H^+A^0/H^0$ processes may already be visible in the 8 TeV LHC data. 

There are numerous and diverse studies of  $SU(2)_L \times SU(2)_R \times U(1)_{B-L}$ models, spanning four decades \cite{Mohapatra:1974hk}.
An interesting aspect of the left-right symmetric models is that they can be embedded in the minimal $SO(10)$ grand unified theory.
This scenario must be significantly modified due to the presence of Dirac masses for right-handed 
neutrinos
required by the CMS $e^+e^-jj$ events.\footnote{Unless the right-handed neutrinos have TeV-scale masses with the split between 
two of them at the MeV scale \cite{Gluza:2015goa}.}
The theory introduced in \cite{Dobrescu:2015qna} involves at least one vectorlike fermion transforming as a doublet under $SU(2)_R $.
This may be part of an additional  $SO(10)$ multiplet, but it may also be associated with entirely different UV completions.

In Section 2 we write down the Higgs potential and analyze its implications for the scalar spectrum. 
In Section 3 we derive the 
interactions of the $W'$ boson with scalars, and compute all $W'$ branching fractions.
The couplings of heavy Higgs bosons to quarks are discussed in Section 4. LHC signals of 
heavy scalars produced in $W'$ decays are the subject of Section 5.
Our conclusions are summarized in Section 6.

\section{Extended Higgs sector }\setcounter{equation}{0}

The Higgs sector of the $SU(2)_L \times SU(2)_R \times U(1)_{B-L}$ gauge theory discussed in \cite{Dobrescu:2015qna} 
consists of two complex scalar fields: an $SU(2)_R$ triplet $T$ of $B-L$ charge +2,  
and an $SU(2)_L \times SU(2)_R$ bidoublet $\Sigma$ of $B-L$ charge 0. 

\subsection{Scalar potential}

The renormalizable Higgs potential is given by
\be
 V(T) + V(T,\Sigma) + V(\Sigma)   ~~,
\ee
where the triplet-only potential is
\be 
V(T)= - M_T^2 \, {\rm Tr} \! \left(T^\dagger T\right)  +
 \frac{\lambda_T}{2}    \left[ {\rm Tr} \! \left(T^\dagger T \right)\right] ^2 
  + \frac{\lambda_T^\prime}{2} \, {\rm Tr} \!  \left[\left(T^\dagger T \right)^2 \right]  ~~.
\ee
The bidoublet-only potential $V(\Sigma)$ is chosen such that by itself it does not generate a VEV for $\Sigma$. This is discussed later, together with 
 $V(T,\Sigma)$, which collects all the terms that involve both scalars.

For $M_T^2 > 0$ and $\lambda_T +  \lambda_T^\prime  > 0$, the $T$ scalar acquires a VEV, which upon an $SU(2)_R \times U(1)_{B-L}$ transformation can be written as
\be
\langle T\rangle = 
 \begin{pmatrix}
 0 & 0  \\
 u_T & 0\\
 \end{pmatrix}   ~~,
 \label{eq:Tvev}
\ee
where the minimization of $V(T)$ gives 
\be
u_T = \frac{M_T}{\sqrt{ \lambda_T + \lambda_T^\prime }} > 0  ~~.
\ee
This breaks $SU(2)_R \times U(1)_{B-L}$ down to the SM hypercharge gauge group, $U(1)_Y$, leading to large masses for the $W'$ and $Z'$ bosons. 
The value of the $T$ VEV is related to the parameters of the $W'$ boson. 
In the next section we will show that the parameters indicated by the LHC mass peaks near 2 TeV imply $u_T  \approx 3\! -\! 4 \; {\rm TeV}$.

The triplet field includes 6 degrees of freedom, and can be written as $T=(T_1,T_2,T_3)$, with $T_i$ $(i = 1,2,3)$ complex scalars, or more explicitly as 
\be
 T = \sum_{i=1}^3  T_i \sigma_i \equiv
 \begin{pmatrix}
 \frac{1}{ \sqrt{2}}  G_R^+  \;  & \;  T^{++}  \\[4mm]
u_T +  \frac{1}{ \sqrt{2}} \left( T^0 + iG^0_R  \right)                  \;  & \;  -  \frac{1}{\sqrt{2}}   G_R^+  \\
 \end{pmatrix}   ~~,
\ee
where $\sigma_i$   are the Pauli matrices, and the factors of $\sqrt{2}$ are required for canonical normalization of the kinetic terms.
The fields of definite electric charge, which are combinations of the $T_i$ components, include three Nambu-Goldstone bosons ($G_R^\pm$, $G_R^0$).
These become the longitudinal degrees of freedom of the $W^{\prime\pm}$ and $Z'$ bosons.
The three remaining fields are a real scalar $T^0$, a doubly-charged scalar $T^{++}$ and its charge conjugate state $T^{--}$.
These have masses given by 
\be 
M_{T^0} = \sqrt{\lambda_T+ \lambda_T^\prime} \; u_T   \;\; \;\;  , \;\;  \;\;  M_{T^{++}} = \sqrt{\lambda_T^\prime}\;  u_T    ~~.
\ee
For quartic couplings in the 0.1--1 range and in the absence of fine tuning, the $T^0$ and $T^{++}$ particles have masses comparable to, or heavier than 
$W'$.

The bidoublet-only potential includes the following terms:  
\bear
 &&  \hspace*{-2.2cm}
 V(\Sigma)  =   M^2_\Sigma \, {\rm Tr} \! \left(\Sigma^\dagger\Sigma\right) 
+ \frac{\lambda_\Sigma}{2}  \left[{\rm Tr} (\Sigma^\dagger\Sigma)\right]^2 
+ \frac{\tilde\lambda_\Sigma}{4} \,  \left| {\rm Tr} \big(\, \widetilde\Sigma^\dagger\Sigma\big) \right|^2
\nonumber \\ [3mm]
&&  \hspace*{-1.1cm}
+  \left[  \frac 1 2 \tilde  M^2_\Sigma \, {\rm Tr} \big(\,\widetilde\Sigma^\dagger\Sigma\big) 
 + \frac{\tilde\lambda_\Sigma^\prime}{8}  \left( {\rm Tr}  \big( \, \widetilde \Sigma^{\dagger}  \Sigma \, \big) \right)^{\! 2} 
+ \frac{\tilde\lambda_\Sigma^{\prime\prime}}{2} \, {\rm Tr} \big(\Sigma^\dagger \Sigma\big) {\rm Tr} \big(\Sigma^\dagger \tilde \Sigma\big)
+ {\rm H.c.}  \right]     ~~,
\label{eq:Vsigma}
\eear
where $ \widetilde{\Sigma}$ is the charge conjugate state of the bidoublet. Other terms, such as  $ {\rm Tr}\! \left[(\Sigma^\dagger \Sigma)^2\right] $
can be written as linear combinations of the terms in Eq.~(\ref{eq:Vsigma}).
We take $M^2_\Sigma , \tilde M^2_\Sigma > 0$, which is a sufficient condition for the minimum of $V(\Sigma)$  to be at $\langle \Sigma \rangle =0$. The mixed terms, which involve both the $T$ and $\Sigma$ scalars, will induce a nonzero VEV.
In terms of fields of definite electric charge, the bidoublet scalar can be written as 
\be
\Sigma =  \left( \begin{array}{cc} \Sigma_1^0 \; & \; \Sigma_2^+ \\  [2mm] \Sigma_1^- \; & \; \Sigma_2^0  \end{array} \right)  ~,
\label{eq:Sigma}
\ee
and its charge conjugate state is
\be
\widetilde{\Sigma} =  \sigma_2 \, \Sigma^* \, \sigma_2  =  \left( \begin{array}{cc}
\Sigma_2^{0*} & -\Sigma_1^+ \\ [2mm]
-\Sigma_2^- & \Sigma_1^{0*}  \end{array} \right)   ~~.
\ee

The mixed terms in the potential are:
\be \hspace*{-0.2cm}
V(T,\Sigma) = \! - \lambda_{T\Sigma}   {\rm Tr} \! \left(T^\dagger T \Sigma^\dagger \Sigma \right) \!
-  \tilde\lambda_{T\Sigma}  {\rm Tr} \big(T^\dagger T \tilde \Sigma^\dagger \widetilde \Sigma \big)\!
 - \!  \left( \! \frac{ \tilde\lambda_{T\Sigma}^\prime }{2} {\rm Tr} \big( T^\dagger T  \big)    {\rm Tr} \big(\Sigma^\dagger  \widetilde\Sigma \big) \!  + {\rm H.c.} \!\!\right) \!
\label{eq:VTS}
\ee
For $\lambda_{T\Sigma} >  M^2_\Sigma/u_T^2$
the VEV of  $T$ induces a negative squared mass for $(\Sigma_2^0 , \Sigma_2^-)$,  given by $- \lambda_{T\Sigma} u_T^2 + M^2_\Sigma < 0 $. 
Likewise, the squared mass for $(\Sigma_1^0 , \Sigma_1^-)$ turns negative, due to the second term in  Eq.~(\ref{eq:VTS}), when
$- \tilde\lambda_{T\Sigma} u_T^2 + M^2_\Sigma < 0 $. In addition, 
a $\Sigma_1^0 \Sigma_2^0$
term is induced for  $\tilde\lambda_{T\Sigma}^\prime \neq 0$. 
Thus, for a range of parameters, the VEV of $\Sigma$ takes the form 
\be
\langle \Sigma \rangle = v_H
 \begin{pmatrix}
\cos\!\beta  & 0  \\
 0 &  e^{i\alpha_{_{\Sigma}}}  \sin\!\beta  \\
 \end{pmatrix}     ~~,
\ee
where $ v_H \simeq 174$ GeV is the electroweak scale. 
We are interested in the case where $u_T /v_H \sim 20$. The effect of the $\Sigma$ VEV   on the  $T^0$ and $T^{++}$ masses
and couplings is thus negligible. At energy scales below the  $T^0$ and $T^{++}$ masses, 
the scalar sector consists only of $\Sigma$, which is the same as two Higgs doublets. 

\bigskip

\subsection{Higgs bosons}

Using the notation of Two-Higgs-Doublet models \cite{Branco:2011iw}, the components of $\Sigma$ defined in Eq.~(\ref{eq:Sigma}) are related to the Higgs 
doublets $H_1$ and $H_2$  as follows:
\bear
&& \hspace*{-1.4cm}
H_1 =  
 \left( \! \begin{array}{c} - \Sigma_1^+ \\ [2mm] {\Sigma_1^{0} }^* \end{array} \! \right)  =
 \left( \begin{array}{c} - H^+ \sin\!\beta  + G^+ \cos\!\beta \\   [4mm] 
  v_H  \cos\!\beta + \frac{1}{\sqrt{2}}  \left(- h^0 \sin\!\alpha  + H^0 \cos\!\alpha  - i A^0 \sin\!\beta + i G^0 \cos\!\beta \right)
    \end{array} \right)    ~,
\nonumber \\ [5mm]
&&  \hspace*{-1.4cm}
H_2 =   \left( \! \begin{array}{c} \Sigma_2^+ \\ [2mm] \Sigma_2^0  \end{array} \! \right)  =
 \left( \! \begin{array}{c}  H^+ \cos\!\beta  + G^+ \sin\!\beta   \\   [4mm] 
 v_H e^{i\alpha_{_{\Sigma}}}  \sin\!\beta + \frac{1}{\sqrt{2}}  \left( h^0 \cos\!\alpha  + H^0 \sin\!\alpha  + i A^0 \cos\!\beta + i G^0 \sin\!\beta \right) 
    \end{array} \! \right)    ,
\label{eq:H1H2}
\eear
where $h^0$ is the SM-like Higgs boson, and $G^\pm, G^0$ are the Nambu-Goldstone bosons, which become the longitudinal $W$ and $Z$. 
We have not included here the effects of the CP-violating phase $\alpha_{_\Sigma}$, which would lead to $H^0-A^0$ mass mixing.
The measurements of  $h^0$ couplings are in good agreement with 
 the SM predictions, implying an alignment limit \cite{Gunion:2002zf}, $\alpha = \beta - \pi/2$.
 In addition, the observed ATLAS events \cite{Aad:2015owa} consistent with the $W' \to WZ$ process
indicate $\sin 2\beta \gtrsim 0.8$ (see Ref.~\cite{Dobrescu:2015qna}), which gives $0.5 \lesssim \tan \!\beta \lesssim 2$.

We have obtained a Two-Higgs-Doublet model with a potential formed of the 
4 quartic terms of $V(\Sigma)$, and 3 independent mass terms from  $V(\Sigma)+V(T,\Sigma)$ with $T$ replaced by its VEV.
The  most general renormalizable potential for two Higgs doublets includes three more quartic terms. 
The allowed potential in our model is a special case of the general Two-Higgs-Doublet model, 
where $\lambda_1 = \lambda_2 = \lambda_3 \equiv \lambda_\Sigma $ 
and $\lambda_6 = \lambda_7 \equiv \tilde \lambda_\Sigma^{\prime\prime}$,  using the standard notation \cite{Branco:2011iw}.
The usual $-m_{12}^2 \widetilde H_1 H_2$ mass mixing term arises here from   the ${\rm Tr} (\widetilde\Sigma^\dagger\Sigma )$
term in $V(\Sigma)$ and the last term in $V(T,\Sigma)$, so that 
\be
m_{12}^2 = {\rm Re} \left( \tilde\lambda_{T\Sigma}^\prime  u_T^2 -  \tilde  M^2_\Sigma  \right) ~~.
\label{eq:m12}
\ee
The squared masses of the charged Higgs boson and of the CP-odd scalar take a simple form:
\bear
&& M_{H^\pm}^2 = \frac{2 m_{12}^2}{  \sin 2\beta} - \left( \tilde\lambda_\Sigma + \tilde\lambda_\Sigma^\prime \right)  v_H^2  ~~,
\nonumber \\ [3mm]
&& M_{A}^2 = M_{H^\pm}^2  + \left( \tilde\lambda_\Sigma - \tilde\lambda_\Sigma^\prime \right)  v_H^2  ~~.
\eear

The value of $m_{12}$ given by Eq.~(\ref{eq:m12}) 
may be comparable to $M_{W'}$, as both are controlled by the $SU(2)_R$ breaking VEV, $u_T$. At the same time,
the weak scale is an order of magnitude smaller than $u_T$, which requires some tuning of the parameters in the potential; a similar tuning
could lead to $m_{12} \ll M_{W'}$. Thus,  theoretically $M_{H^\pm}$ and $ M_{A}$ may be anywhere between the weak scale and $\sim M_{W'}$.

Various searches for the heavy Higgs bosons set mass limits substantially above the mass of the SM-like Higgs boson ($M_h = 125$ GeV).
It is sufficient then to expand in $(v_H/m_{12})^2 \ll 1$.
The heavy CP-even scalar, $H^0$, has a squared mass
\be 
M_{H^0}^2 = M_{H^\pm}^2 + \left( \tilde\lambda_\Sigma +\tilde\lambda_\Sigma^\prime \right)\cos^2 \! 2\beta \; v_H^2  ~~.
\ee
Note that the relative mass splittings between $H^\pm$, $A^0$ and $H^0$ are small. These four states
approximately form a heavy $SU(2)_W$ doublet of zero VEV.

The mass of the SM-like Higgs boson is related to the quartic couplings by
\be
M_h^2 = \left[ \left( \tilde\lambda_\Sigma +\tilde\lambda_\Sigma^\prime \right)\sin^2 \! 2\beta 
+  2 \lambda_\Sigma \rule{0pt}{14pt} \right] v_H^2   ~~.
\ee
This provides an estimate of the typical values of the quartic couplings: 
for $\tan\beta \to 1$, $2 \lambda_\Sigma+\tilde\lambda_\Sigma +\tilde\lambda_\Sigma^\prime \approx 0.5$.
The departure from the alignment limit is given by
\be
\alpha - \beta + \frac \pi 2 
\simeq -\frac 1 2 \left(\tilde \lambda_\Sigma+\tilde \lambda_\Sigma^\prime \right) \, \sin 4\beta \, \frac {v_H^2} {M_{H^\pm}^2} ~~.
\label{eq:alignment}
\ee

\bigskip

\section{$W'$ decays into heavy Higgs bosons}\setcounter{equation}{0}

Given the Higgs sector discussed in the previous section, which breaks the  
$SU(2)_L \times SU(2)_R \times U(1)_{B-L}$ gauge symmetry down to the $U(1)_{\rm em}$ group of electromagnetism,
we can now derive the $W'$ couplings to bosons and the ensuing decay widths.

\subsection{Couplings of the $W'$ to bosons}

The $SU(2)_L$ and $SU(2)_R$ gauge bosons include electrically-charged states, $W_L^{\pm \mu}$ and $W_R^{\pm \mu}$, respectively. 
The kinetic terms for the $T$ and $\Sigma$ scalars give rise to the following mass terms for the charged gauge boson  
\be
 \left( W_L^{+ \mu} , W_R^{+ \mu} \right)   \left( \begin{array}{cc}  
\;\; \;\; {\displaystyle   g_{2_L}^2 \frac{v_H^2 }{ 2} }                            \;  &  \;  -  g_{2_L} g_{2_R} \,    {\displaystyle  \frac{v_H^2}{ 2} } \, \sin 2\beta  \\ [5mm]
  -  g_{2_L} g_{2_R} \,   {\displaystyle  \frac{v_H^2}{ 2} } \,  \sin 2\beta  \; &  \;   g_{2_R}^2 \left( u_T^2 +  {\displaystyle  \frac{v_H^2}{ 2} } \right)   \end{array} \right) 
  \left( \begin{array}{c}    W_{L \mu}^-  \\ [4mm] W_{R \mu}^-   \end{array} \right)   ~~,
\label{eq:mwp}
\ee
where $g_{2_L}$ and $g_{2_R}$ are the $SU(2)_L\times SU(2)_R$ gauge couplings.
Since we are interested in the case $M_{W'} \gg M_{W}$, we diagonalize the above mass matrix 
by keeping only the leading term in $(M_W/M_{W'})^2$. 
All equations that follow are valid up to corrections of order $(M_W/M_{W'})^2 \approx 0.2\%$, where we used $M_{W'} \approx 1.9$ TeV.
The $SU(2)_L$ gauge coupling is given by the SM $SU(2)_W$ gauge coupling $g$,
while the $SU(2)_R$ gauge coupling is given by the $W'$ coupling $g_{_{\rm R}}$  to the $u\bar d$  quarks in the gauge eigenstate basis:
\bear
&& g_{2_L} = g \approx 0.65  ~~,
\nonumber \\ [3mm]
&& g_{2_R} = g_{_{\rm R}} \approx 0.45\!-\!0.6  ~~,
\eear 
where the range for $g_{_{\rm R}}$ is required in order to explain \cite{Dobrescu:2015qna} the LHC mass peaks near 2 TeV.

The physical bosons, $W$ and $W'$, are admixtures of $W_L^{\pm \mu}$ and $W_R^{\pm \mu}$:
\bear
W^\pm_\mu & = & W_{L\, \mu}^\pm \cos\theta_{\!_+} + W_{R\, \mu}^\pm \sin\theta_{\!_+}  ~~,
\nonumber \\ [3mm]
W^{\prime \pm}_\mu & = & \!\! -W_{L\, \mu}^\pm \sin\theta_{\!_+} + W_{R\, \mu}^\pm \cos\theta_{\!_+}  ~~.
\eear
The $W_L^{\pm \mu}-W_R^{\pm \mu}$ mixing angle $\theta_{\!_+}$ satisfies
\be
\sin\theta_{\!_+} = \frac{g_{_{\rm R}}}{g } \left( \frac{ M_W}{ M_{W'}}\right)^{\! 2} \sin 2\beta  ~~,
\ee
and the $W$ and $W'$ masses are given by 
\bear
M_W =  \frac{ g \,  v_H }{ \sqrt{2} }  ~~,
\nonumber \\ [3mm]
M_{W'} = g_{_{\rm R}} \,  u_T  ~~.
\eear
For $M_{W'} \approx 1.9$ TeV and $g_{_{\rm R}} \! \approx 0.45$--0.6 (as determined in \cite{Dobrescu:2015qna}, by comparing the 
$W'$ production cross section to the CMS dijet excess \cite{Khachatryan:2015sja}), we find 
the $SU(2)_R$ breaking scale $u_T \approx 3$--4 TeV. This set of parameters is compatible with constraints from electroweak precision observable~\cite{Langacker:1989xa}.

The $W^\prime W Z$ interactions are given by 
\be
 \frac {g}  {c_W} \sin\theta_{\!_+} \,  i \left[ W^{\prime + }_\mu \left(W^-_\nu \partial^{[\nu}Z^{\mu]} +  Z_\nu \partial^{[\mu} W^{-\nu]}  \rule{0pt}{12pt}  \right)
+Z_\nu W^-_\mu \partial^{[\nu} W^{\prime +\mu]}   \rule{0pt}{14pt} \right]  +  {\rm H.c.}
\label{eq:WWZ}
\ee
Here $c_W \equiv \cos\theta_W$ is the usual SM parameter, 
and $[\mu,~\nu]$ represents commutation ($\mu\nu-\nu\mu$) of the two Lorentz indices. Both $W^+_L W^-_L Z$ and $W^+_R W^-_R Z$ terms contribute to the above $W^\prime W Z$ interactions through $W_L$-$W_R$ mixing, with coupling strengths $g c_W$ and $g s_W^2/c_W$, respectively.
Note that the coefficient of Eq.~(\ref{eq:WWZ}) can also be written as $g_{_{\rm R}} (M_W/M_{W'})^2 \sin 2\beta  /c_W$. Comparing this form with Eq.~(4) of 
\cite{Dobrescu:2015qna} gives $\xi_Z = \sin 2\beta $.

The $W^\prime$ interactions with a $W$ and a neutral Higgs boson arise from the $\Sigma$ kinetic term:
\be
- g_{_{\rm R}} M_W   \; W^{\prime +}_\mu \, W^{\mu -} \left[ h^0  \cos (\alpha + \beta) + H^0 \sin  (\alpha + \beta) - i  A^0  \cos 2\beta \right]  +  {\rm H.c.} 
\label{eq:WWH}
\ee
also a $W^{\prime\pm}$ interaction with $Z H^\mp$, which gets contributions from the kinetic terms of $\Sigma$ and $T$:
\be
- \frac {g_{_{\rm R}}} {c_W} M_W \cos 2\beta  \; \, W^{\prime\pm}_\mu \, Z^\mu H^\mp  ~~.
\ee

The $W^{\prime \pm}$ couplings to $H^\mp$ and one of the neutral Higgs bosons are the following:
\be
\frac{ i g_{_{\rm R}} }{2} \, W^{\prime +}_\mu \!  \left[  - \cos ( \alpha \! + \! \beta )\,  \left( H^-  \partial^\mu  H^0 - H^0  \partial^\mu  H^- \right)
+ \sin (  \alpha \! + \! \beta )\,  \left( H^-  \partial^\mu  h^0 - h^0  \partial^\mu  H^- \right) \rule{0pt}{13pt} \!\right] +  {\rm H.c.}
\ee
In the alignment limit with $\tan\beta \to 1$ the above $W^{\prime \pm} H^\mp h^0 $ coupling vanishes, while the  $W'^\pm H^\mp H^0 $ coupling reaches its maximum.
Finally, the $W^{\prime \pm} H^{\mp} A^0 $ coupling is
\be
\frac{g_{_{\rm R}} }{2} \sin2\beta \,\, W^{\prime +}_\mu  \left( H^-  \partial^\mu  A^0 - A^0  \partial^\mu  H^- \right) +  {\rm H.c.}
\ee

\bigskip

\subsection{$W'$ branching fractions}

The dominant decay modes of $W^\prime$ are into fermion pairs, mainly quark-antiquark pairs.
$W^\prime$ can also decay into bosons, leading to challenging and interesting phenomenology. 
Amongst many possible bosonic decay modes, $W^\prime \to WZ$ and $W^\prime \to W h^0$ are inevitable, originating from the kinetic term of the bidoublet scalar field $\Sigma$. Unlike other bosonic decays discussed below, the widths of these two modes do not depend on the unknown masses of 
heavy scalars. Eqs.~(\ref{eq:WWZ}) and (\ref{eq:WWH}) imply
\bear
&& \Gamma(W^\prime \to WZ) \simeq \frac {g_{_{\rm R}}^2} {192\pi}   \sin^2 \! 2\beta \; M_{W^\prime}  ~~,
\nonumber \\ [3mm]
&& \Gamma(W^\prime \to Wh^0) \simeq \frac {g_{_{\rm R}}^2} {192\pi} \cos^2(\alpha+\beta) \; M_{W^\prime} ~~,
\label{eq:WZWh}
\eear
where terms of order $(M_{W,Z,h}/M_{W^\prime})^2$ have been neglected, and the approximate relation $c_W M_Z \simeq M_W$ has been used. We use full expressions for numerical study later in this section, which lead to a slight difference between the partial widths of these two $W^\prime$ decay modes, arising from differences in underlying dynamics and kinematics.

The equivalence theorem requires that $W'$ decays into fields that are part of the 
same Higgs doublet (the longitudinal $Z$ and $h^0$  for $\alpha \to \beta - \pi/2$ in this case) 
have equal decay widths up to electroweak symmetry breaking effects and phase-space factors. 
As mentioned in Ref.~\cite{Dobrescu:2015qna},
in the alignment limit $\Gamma(W^\prime \to WZ) \simeq \Gamma(W^\prime \to Wh^0)$.
 
Besides $h^0$ and the longitudinal $W$ and $Z$, the bidoublet field $\Sigma$ includes the heavy scalars $H^\pm, H^0, A^0$.
 The range of allowed masses for these particles spans more than an order of magnitude, from 
 the weak scale to the $SU(2)_R$ breaking scale.  If they are lighter than the $W'$ boson, then  
 $W^\prime$ decays may provide the main mechanisms for production of these scalar particles at hadron colliders.
 The  $W^\prime$ decays into a heavy scalar and an electroweak boson have widths 
\bear
&& \Gamma(W^{\prime\pm} \to Z H^\pm) \simeq \frac {g_{_{\rm R}}^2} {192\pi}  \cos^2\! 2\beta \; 
M_{W^\prime}\left(1-\frac {M_{H^\pm}^2} {M_{W^\prime}^2}\right)^{\!3}  ~~,
\nonumber \\ [3mm]
&& \Gamma(W^{\prime} \to W \phi) \simeq \frac {g_{_{\rm R}}^2} {192\pi} \,  \xi^2_{\phi} \, M_{W^\prime}\left(1-\frac {M_\phi^2} {M_{W^\prime}^2}\right)^{\!3},
\label{eq:Wphi}
\eear 
where $\phi$ labels the heavy neutral scalars, and $\xi_{\phi}$ follows from the couplings in Eq.~(\ref{eq:WWH}): 
\begin{equation}
\xi_{\phi}^2 =\left\{\ba{lcl}
\sin^2(\alpha+\beta) &,&~{\rm for}~\phi=H^0  ~~,\\ [2mm]
\cos^2 \! 2\beta &,&~{\rm for}~\phi=A^0  ~~.
\ea\right.
\label{eq:xiphi}
\end{equation}
Here we neglected terms of order  $M_W/M_{W^\prime}$, which are relevant only for $M_{H^\pm} + M_W$ close to $M_{W^\prime}$. 
The exact expressions are given by replacing the last factor in Eq.~(\ref{eq:Wphi}), 
\be
\left(1-\frac {M_\phi^2} {M_{W^\prime}^2}\right)^{\!3} \to 
F\!\left( \frac {M_\phi^2} {M_{W^\prime}^2} , \frac {M_W^2} {M_{W^\prime}^2} \right)  ~~,
\ee
and a similar replacement for $W^{\prime\pm} \to Z H^\pm$,
where the function $F$ is defined by 
\be
F(x,y) = f^3 (x,y) + 8y (1+ 2xy) \, f(x,y)
\ee
and $f$ is the phase-space suppression factor:
\be
 f(x,y) = \left( 1 - 2(x+y) + (x-y)^2 \rule{0pt}{12pt} \right)^{1/2}   ~~.
\ee

Similarly, the $W^\prime$ decay width into a charged Higgs boson and the SM-like Higgs particle is
\begin{equation}
\Gamma(W^{\prime\pm} \to H^\pm h^0)=\frac {g_{_{\rm R}}^2 } {192\pi} \sin^2(\alpha+\beta) \, M_{W^\prime} \left(1-\frac {M_{H^\pm}^2} {M_{W^\prime}^2}\right)^{\!3}  ~~,
\end{equation}
The terms of order $M_h/M_{W^\prime}$, neglected here, are taken into account by replacing the last factor in the above equation
by $f^3(M_{H^\pm}^2/M_{W^\prime}^2 \, , M_h^2/M_{W^\prime}^2)$.
For $M_{H^\pm} < M_{W'}/2$,  the $W^\prime$ can also decay into a pair of heavy scalars, with widths 
\begin{equation}
\Gamma(W^{\prime\pm} \to  H^\pm \phi)=\frac {g_{_{\rm R}}^2 } {192\pi} \left(1 - \xi_\phi^2 \right) \, M_{W^\prime} \left(1-\frac {4 M_{H^\pm}^2} {M_{W^\prime}^2}\right)^{\!3/2} ~~,
\end{equation}
where $\xi_\phi^2$ is defined in Eq.~(\ref{eq:xiphi}).
Here we used the  $M_A = M_{H^0} = M_{H^\pm}$ limit; the exact expression is obtained by replacing the last factor with 
$f^3(M_{H^\pm}^2/M_{W^\prime}^2 \, , M_\phi^2/M_{W^\prime}^2)$.

The 8 widths for $W'$ decays into bosons shown in this section satisfy the equivalence theorem 
for $M_{H^\pm}^2 \ll M_{W'}^2$ in the alignment limit.
Summing over these 8 widths gives 
\be
\Gamma (W^{\prime} \to {\rm bosons}) = \frac {g_{_{\rm R}}^2 } {48\pi}  M_{W^\prime} \left[1 - O(4M_{H^\pm}^2/M_{W^\prime}^2)\right]~~.
\label{eq:sum}
\ee
Note that the leading order in $M_{H^\pm}^2/M_{W^\prime}^2$ is independent of $\tan\beta$.

The dominant decay modes of $W^\prime$ are into quarks, and  have the following widths:
\begin{equation}
\Gamma(W^\prime \to jj) = \frac {g_R^2} {8\pi} \, M_{W^\prime}
\end{equation}
for light flavors, and 
\begin{equation}
\Gamma(W^{\prime +} \to t \bar b) = \frac {g_R^2} {16\pi} M_{W^\prime} \left(1+\frac {m_t^2} {2 M_{W^\prime}^2}\right) \left(1-\frac {m_t^2} {M_{W^\prime}^2}\right)^{\!2}
\end{equation}
for heavy flavors. QCD corrections increase these two widths by about 3\%.
The decay widths into a $\tau$  lepton or an electron and the $N^\tau$ right-handed neutrino are
\bear
&& \Gamma(W^\prime \to \tau N^\tau) = \frac {g_{_{\rm R}}^2} {48\pi} (1 - s_{\theta_\ell}^2 ) M_{W^\prime} \left(1+\frac {m_{N^\tau}^2} {2 M_{W^\prime}^2}\right) \left(1-\frac {m_{N^\tau}^2} {M_{W^\prime}^2}\right)^{\!2}  ~~,
\nonumber \\ [3mm]
&& \Gamma(W^\prime \to e N^\tau) = \frac{s_{\theta_\ell}^2}{1 - s_{\theta_\ell}^2}\,  \Gamma(W^\prime \to \tau N^\tau)  ~~,
\label{eq:eN}
\eear
where $s_{\theta_e}$ is the coefficient of the $g_{_{\rm R}} W'_\nu \, \bar e_R \gamma^\nu N^\tau_R$ interaction term in the Lagrangian,
and the analogous coefficient for the muon satisfies $s_{\theta_\mu} \ll s_{\theta_e}$.
The baseline $W'$ model used in \cite{Dobrescu:2015qna} to explain the excess events near 2 TeV reported in several channels by ATLAS and CMS, including $e^+e^-jj$ \cite{Khachatryan:2014dka},  has  $s_{\theta_e} \approx 0.5$, and is consistent with all current constraints on flavor-changing 
processes.\footnote{The tension with the $\mu \to e \gamma$ limit mentioned in \cite{Brehmer:2015cia} is not a concern for our model  given that $s_{\theta_\mu}\! \to 0$ is a natural limit.} 
We emphasize that taking into account the $e^+e^-jj$ excess was crucial in identifying  our baseline $W'$ model 
(without it, leptophobic $W'$ models \cite{Gao:2015irw} are interesting alternatives). If the  $W'$ boson will be discovered in Run 2, then
the $eejj$ process will allow various tests of the underlying couplings \cite{Han:2012vk}.

\begin{figure}[t]
\centering
\includegraphics[width=0.7\textwidth]{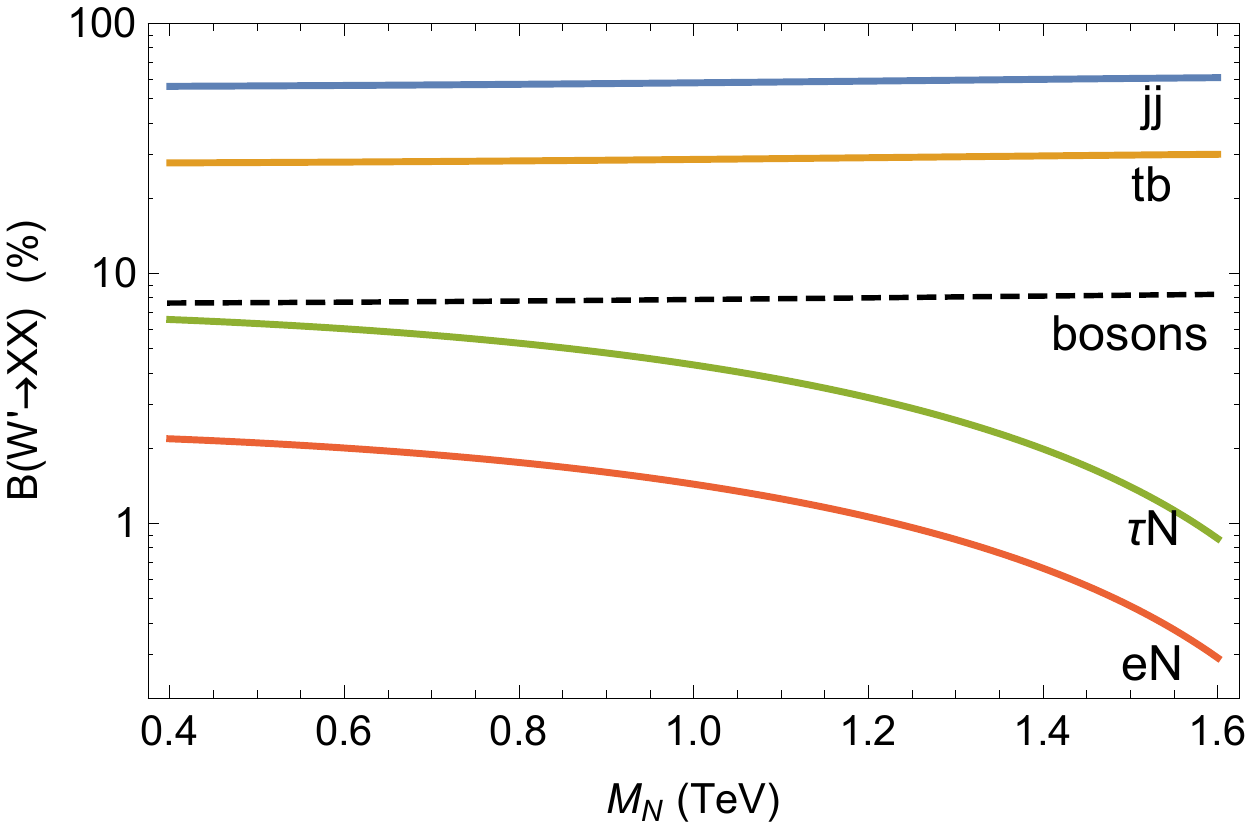} 
\caption[]{Branching fractions of $W^\prime$ for $M_{W'}= 1.9$ TeV and $A^0$, $H^0$, $H^\pm$ masses of 500 GeV. 
The dashed line represents the sum of all 8 bosonic decay modes. The $W'\to\tau N^\tau$ and $e N^\tau$ widths are computed for 
$s_{\theta_e} = 0.5$.  }\label{fig:wpbr}
\end{figure}

The $W^\prime \to jj$ and $W^\prime \to tb$ channels have a combined branching fraction of approximately $86\%$,
as shown in Figure~\ref{fig:wpbr}. 
The branching fractions for  $W' \to \tau N_R^\tau$ and $W' \to e N_R^\tau$ add up to $6\%$ for the benchmark value of $M_{N_R^\tau} = 1$ TeV. 
The remaining decay modes are into bosons, which together have a branching fraction of approximately 8\% for $M_{H^\pm } = 500$ GeV, as indicated by the dashed 
line in Figure~\ref{fig:wpbr}. 
Note that the difference between  $\tan\beta = 1$ and 2 cannot be resolved, as expected from Eq.~(\ref{eq:sum}).

\begin{figure}[t]
\centering
\includegraphics[width=0.485\textwidth]{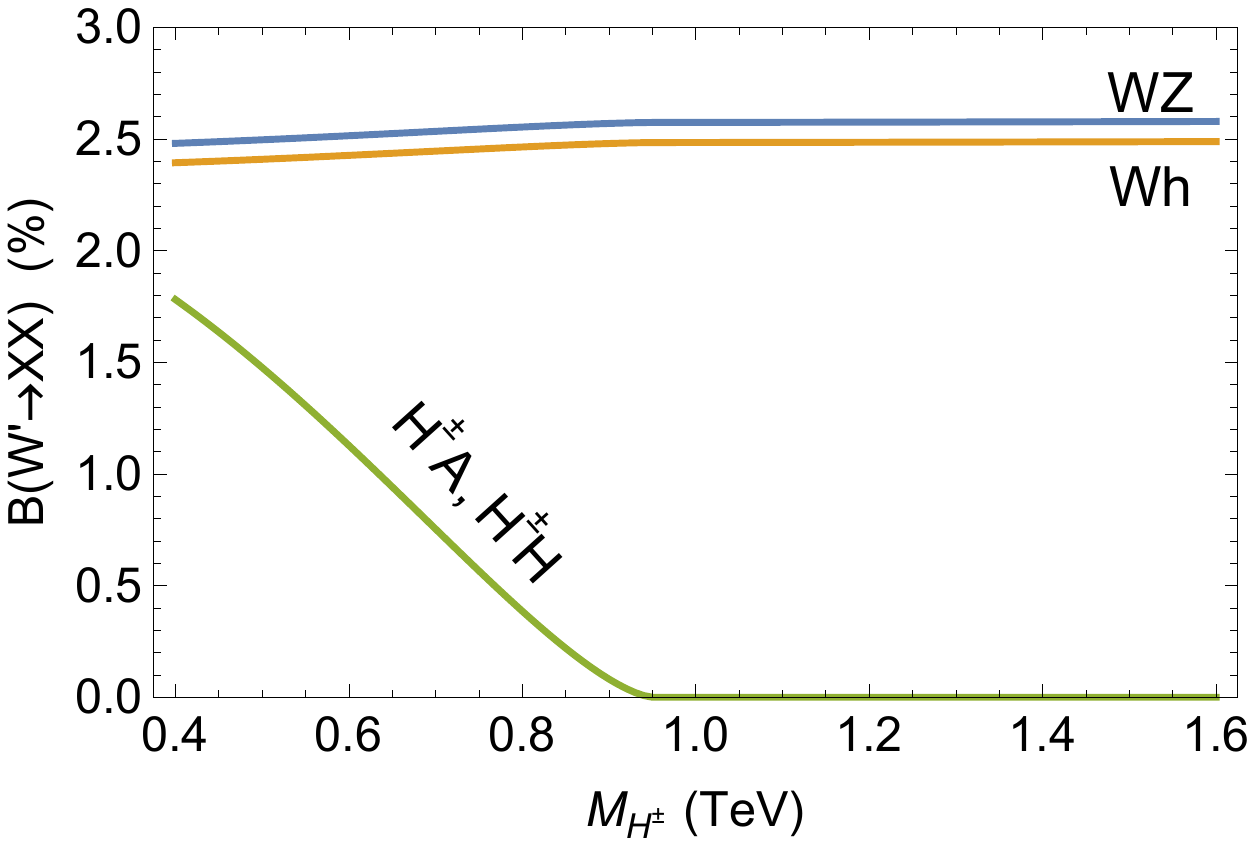}  
\includegraphics[width=0.485\textwidth]{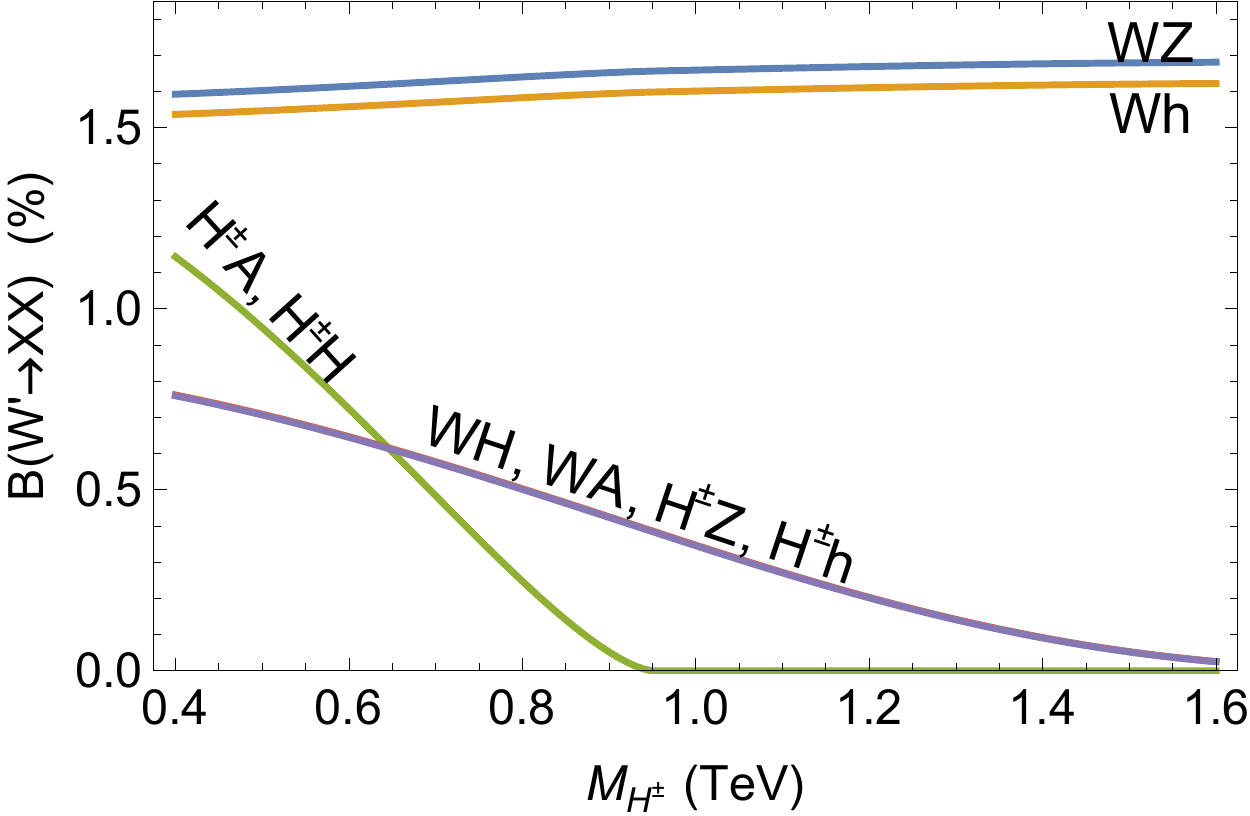}  
\caption[]{Branching fractions for $W^\prime$ decays into bosons,  for  
$\tan\beta =1 $ (left panel) and $\tan\beta =2$ (right panel).
The mass of the $N^\tau$ right-handed neutrino is fixed at 1 TeV, and  $M_{W'}= 1.9$ TeV. 
The $H^\pm A^0$ and $H^\pm H^0$ branching fractions are equal and not summed here (similarly for $WA^0$, $WH^0$, $H^\pm Z$ and $H^\pm h^0$).  
}\label{fig:wpbr-bosons}
\end{figure}

Eqs.~(\ref{eq:WZWh})-(\ref{eq:eN}) show that the $W'$ widths into bosons depend on the almost degenerate
masses of the heavy scalars, and on $\sin 2\beta$ whose value can be between 0.8 and 1, 
as favored by the CMS and ATLAS mass peaks near $M_{W^\prime} = 1.9$~TeV discussed in~\cite{Dobrescu:2015qna}. 
The dependence on the CP-even Higgs mixing angle $\alpha$ is very weak, due to the alignment limit 
Eq.~(\ref{eq:alignment}), which implies $\cos(\alpha+\beta) \simeq \sin2\beta$ and $\cos(\alpha-\beta) \ll 1$. 
We show the $W^\prime$ branching fractions in  Figure~\ref{fig:wpbr-bosons} as a function of $M_{H^\pm} \simeq M_{A} \simeq M_{H^0}$,
for $\sin 2\beta = 1$ ({\it i.e}, $\tan\beta = 1$) and $\sin 2\beta = 0.8$ ({\it i.e}, $\tan\beta = 2$ or 1/2).
The $W^\prime$ branching fractions to $Wh^0$ and $WZ$ are almost constant 
because these partial widths do not depend on the masses of the heavy scalars. The branching fractions of $W^\prime$ into $H^\pm H^0$ and $H^\pm A^0$ are equal because of the alignment condition and the approximate mass degeneracy.
Similarly, decay widths of $WH^0$, $WA^0$, $H^\pm Z$ and $H^\pm h^0$ are equal, suppressed 
by $\cos^2 2\beta$, and with a different phase-space function than the two-heavy-scalar modes. 
We can see from Figure~\ref{fig:wpbr-bosons} that the $\tan\beta=1$ case is simple: $WZ$, $Wh^0$, $H^+ H^0$ and $H^+ A^0$  modes are maximal, and the other decay modes vanish.
When $\tan\beta$ deviates from unity, the other four decay modes start to emerge with sub-percent level branching fractions.

\bigskip

\section{Quark masses} \setcounter{equation}{0}

The $SU(2)_L\times SU(2)_R \times U(1)_{B-L}$ gauge symmetry allows quark masses to be generated by
Yukawa couplings to the bidoublet $\Sigma$: 
\be
- \overline Q_L^i \left( y_{ij}  \Sigma + \tilde y_{ij}  \widetilde \Sigma \right) Q_R^j  + {\rm H.c.} 
\ee
Here $i,j=1,2,3$ label the SM fermion generations, $Q_L^i=(u_L^i , d_L^i)$ is the $SU(2)_L$
quark doublet of the $i$th generation, and $Q_R^i=(u_R^i , d_R^i)$ is the corresponding $SU(2)_R$ doublet.
The Yukawa couplings $y_{ij}$ and $\tilde y_{ij}$ are complex numbers.
The mass terms for the up- and down-type quarks are then 
\be 
- v_H \, \overline u_L^i \left( y_{ij}  \cos\!\beta  +  \tilde y_{ij} \sin\!\beta \right) u_R^j 
- v_H \, \overline d_L^i \left( y_{ij}  \sin\!\beta  +  \tilde y_{ij} \cos\!\beta \right) d_R^j  + {\rm H.c.} 
\label{eq:quark-mass-terms}
\ee
These terms highlight a problem. 
As we are interested in $\tan\beta = O(1)$, the above terms generically induce
masses of the same order of magnitude for up- and down-type quarks.
Even though the fermion mass hierarchies are not understood in the SM, they can be fitted by 
appropriately small Yukawa couplings. 
In the case of Eq.~(\ref{eq:quark-mass-terms}), the known ratio of the bottom and top quark masses can be 
fitted only by tuning various parameters. For example, $m_b \ll v_H$ can be achieved by imposing
$\tilde y_{33} / y_{33} \simeq - \tan\!\beta$. In the absence of an explanation for the tuning of these 
independent parameters, it is useful to explore alternative mechanisms for fermion mass generations.

Consider the case where $y_{ij}$  and  $\tilde y_{ij}$ are negligibly small. 
A different type of gauge-invariant operator that can generate the quark masses is 
\be
- \frac{C_{ij}}{u_T^2} \; \, \overline Q_L^i \, \widetilde \Sigma \,
 {T^\dagger T}  \,  Q_R^j  + {\rm H.c.}  ~~,
 \label{eq:sigmaTT}
\ee
where the $SU(2)_R$ indices of the scalar triplet $T$ are contracted with 
$T^\dagger$ such that $T^\dagger T$ belongs to the triplet representation of $SU(2)_R$. 
The flavor-dependent coefficients $C_{ij}$ are complex dimensionless parameters, and form a $3\times 3$ mass matrix $C$.
We assume that the analogous operator with $\widetilde \Sigma$ replaced by $\Sigma$ 
has negligibly small coefficients. 
Replacing $T$ by its VEV given in Eq.~(\ref{eq:Tvev}), the dimension-6 operator of Eq.~(\ref{eq:sigmaTT}) 
generates a $3\times 3$ mass matrix for the up-type quarks,
\be 
{\cal M}_u =   v_H \sin\!\beta  \; C ~~~,
\ee
while the down-type quarks remain massless at this stage.

There is, however, an analogous effective operator
with $T$ replaced by $\widetilde T \equiv \sigma_2 T^* \sigma_2$, 
\be
- \frac{\tilde C_{ij}}{u_T^2} \; \, \overline Q_L^i \, \widetilde \Sigma \,
 {\widetilde T^\dagger \widetilde T}  \,  Q_R^j  + {\rm H.c.} 
 \label{eq:sigmaTTtilde}
\ee
This generates a mass matrix only for down-type quarks:
\be 
{\cal M}_d = v_H \cos\!\beta  \; \tilde C  ~~.
\ee
Given that the coefficients $\tilde C_{ij}$ are different than $C_{ij}$, we obtain
mass matrices for the up- and down-type quarks that are independent of each other,
as in the SM. 
Various possibilities for the origin of the effective operators (\ref{eq:sigmaTT}) and (\ref{eq:sigmaTTtilde})
will be discussed elsewhere.

The operators (\ref{eq:sigmaTT}) and (\ref{eq:sigmaTTtilde}) have an additional 
useful property: they separate the contributions of the two Higgs doublets to the quark masses.
Notice that the embedding (\ref{eq:H1H2}) of $H_1$ and $H_2$ in $\Sigma$ 
implies $\widetilde\Sigma = (\widetilde H_2, H_1)$, where $\widetilde H_2 = i \sigma_2 H_2^*$.
From Eqs.~(\ref{eq:sigmaTT}) and (\ref{eq:sigmaTTtilde}) then follows
\be
- C_{ij} \; \, \overline Q_L^i \, \widetilde H_2  \,  u_R^j 
- \tilde C_{ij} \; \, \overline Q_L^i \, H_1 \,  d_R^j 
 + {\rm H.c.}  
 \label{eq:2HDM}
\ee
We have obtained the Yukawa couplings of the Two-Higgs-doublet model of Type II. This automatically 
avoids tree-level flavor-changing processes (see~\cite{Blanke:2011ry} and references therein). The couplings of the heavy Higgs bosons to the physical eigenstates of
the quarks are proportional to the quark masses, with overall coefficients of $\sin\!\beta$ for up-type quarks
and $\cos\!\beta$ for down-type quarks. 
For $0.5 \lesssim \tan\!\beta \lesssim 2$ and close to the alignment limit, as indicated \cite{Dobrescu:2015qna}
by the $W'$ signals near 2 TeV, the branching fractions for $H^+ \to t \bar b$ and $H^0, A^0 \to t \bar t$ 
are almost 100\%.

\bigskip

\section{Signals of heavy scalars produced in $W'$ decays}\setcounter{equation}{0}

Even though the $W'$ decays into heavy scalars have only percent-level branching fractions (see Figure 2), the large $W'$ production cross 
section at the LHC \cite{Dobrescu:2015qna}
(200--350 fb at $\sqrt{s} = 8$ TeV, and 1--2 pb at $\sqrt{s} = 13$ TeV) makes these decays promising discovery channels.
The $pp \to W' \to H^\pm A^0$ and  $H^\pm H^0$ processes lead to cascade decays that end up with $3W+4b$ final states, including
\be 
pp \to W^{\prime +} \to H^+ A^0/H^0 \to t \, \bar b \,  t \, \bar t  \to W^+W^+ W^- \! +4b \to 
\left\{ \ba{c} \ell^+ \ell^+jj  + 4b + \met \\ [1mm]
\;\; {\rm or} \;\; \\  [1mm]
\ell^+ \ell^+ \ell^-\! +4b + \met   
\ea  \right. 
\label{eq:3tb}
\ee
where $\ell = e$ or $\mu$. The charge conjugate processes lead to a number of events smaller by a factor of almost 2, due to the 
smaller $W^-$ production in $pp$ collisions.
The cross section times branching fraction relevant for these processes is shown in Figure 3 for the LHC at 8 and 13 TeV.
We used there $\tan\beta =1$ and a $W'$ production cross section of 300 fb at $\sqrt{s} = 8$ TeV for $M_{W'} = 1.9$ TeV (implying
a cross section of 1.7 pb at $\sqrt{s} = 13$ TeV).

An additional $W'$ decay mode that contributes to final states with same-sign leptons or three leptons and $b$ jets is
\be
pp \to  W^{\prime +} \to \tau^+  N^\tau \to \tau^+ \tau^- t \bar b \to \ell^+ \nu \bar \nu \, \tau^-   W^+ b\bar b
\to 
\left\{ \ba{c} \ell^+ \ell^+ \tau^-_h b \bar b + \met \\ [1mm]
\;\; {\rm or} \;\; \\  [1mm]
\ell^+ \ell^+ \ell^- b \bar b + \met   
\ea  \right. 
\ee
The branching fraction for the $N^\tau$ decay used here is  $B(N^\tau \to \tau^- t \bar b) \approx 23\%$, while 
similar final states arise from $B(N^\tau \to e^- t \bar b) \approx 8\%$.
The cascade decays  $W^{\prime +} \to e^+  N^\tau$ with $N^\tau \to \tau t \bar b$ or $e t \bar b$ (the latter process is mentioned in 
\cite{Aguilar-Saavedra:2014ola}) also contribute.

The heavy scalars may also be produced directly, without $W'$ decays. A promising channel is the production of $A^0$ and $H^0$
in association with a $t \bar t$ pair \cite{Craig:2015jba}:
\be
pp \to  t \bar t A^0/H^0 \to 4t \to 4W+4b  ~~.
\label{eq:4t}
\ee
Compared to the SM $t\bar t h^0$ process at large $M_h$, each of the above two processes has a cross section 
scaled by $(\tan \beta)^{-2}$.
In Figure 3 we show the cross section for this process, computed 
at leading order using MadGraph5\_aMC@NLO~\cite{Alwall:2014hca} 
with the MSTWnlo2008~\cite{Martin:2009iq} parton distribution functions. 
We see that  the processes listed in Eqs.~(\ref{eq:3tb}) and (\ref{eq:4t}) have comparable 
cross section times branching fractions.

\begin{figure}[t]
\centering
\includegraphics[width=0.69\textwidth]{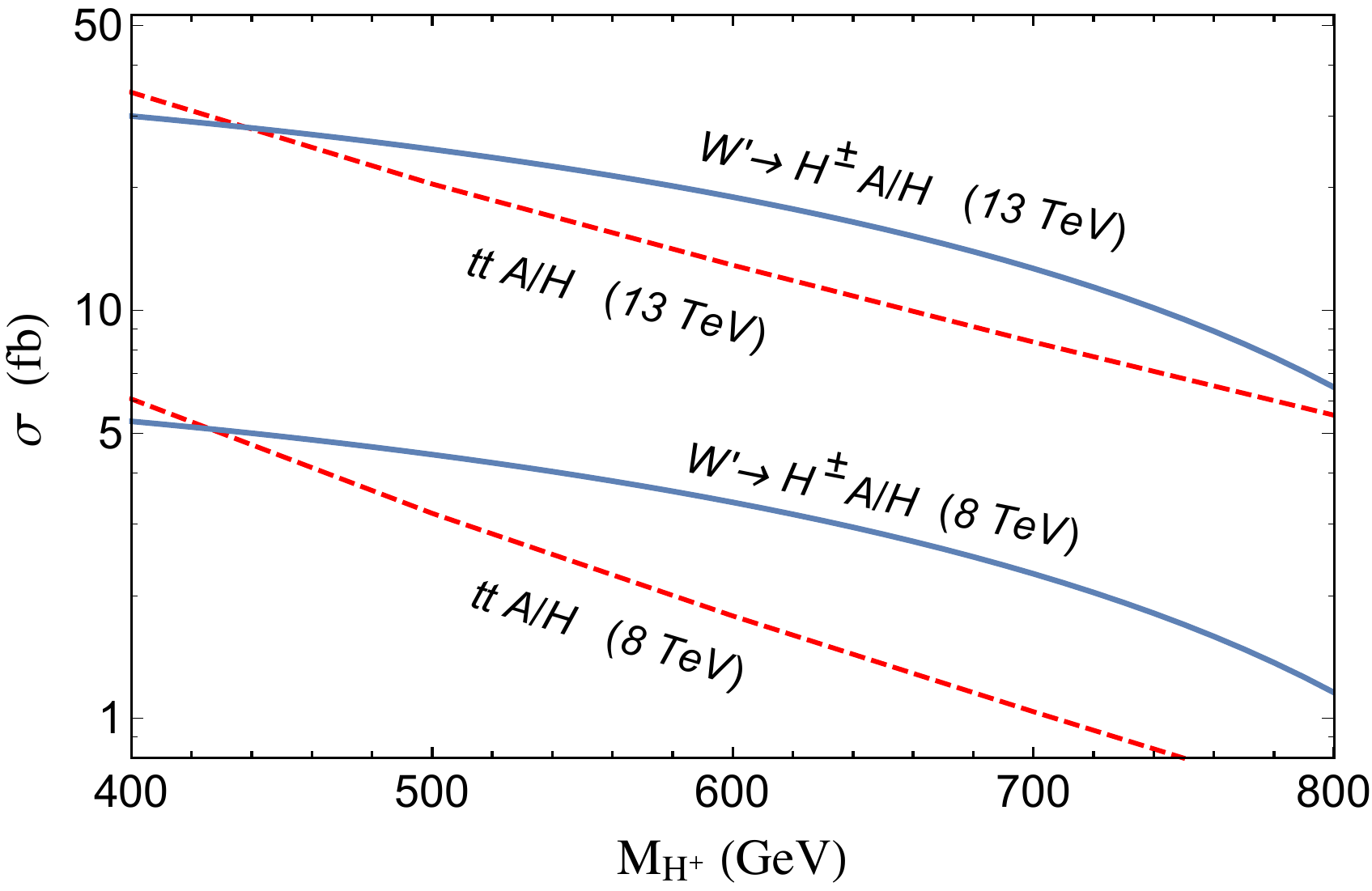} 
\caption[]{Production cross section times branching fractions for the sum of $pp \to W' \to H^\pm A^0 , H^\pm H^0 $ (solid lines),
and for the sum of $pp \to t \bar t A^0 , t \bar t H^0$ (dashed lines), at $\sqrt {s} = 8$ and 13 TeV.
The parameters used here are  $\tan\beta = 1$, $M_{W'} = 1.9$ TeV  and a $W'$ production cross section of 300 fb at 8 TeV.}\label{fig:AH}
\end{figure}

The ATLAS search \cite{Aad:2015gdg} for same-sign leptons and $b$-jets shows some deviations from the SM
in two signal regions designed for four-top final states, with a pair of same-sign leptons and $H_T > 700$ GeV.
In the signal region ``SR4t3" (exactly two $b$-jets and $\met > 100$ GeV)  there are 12 events
with an expected background of $4.3 \pm 1.1 \pm 1.1$ events. 
In the  signal region ``SR4t4" (3 or more $b$-jets) the search found 6 events with an expected background of $1.1 \pm 0.9 \pm 0.4$ events. 
We estimate that the combination of these signals represents a $\sim \!3\sigma$ excess over the SM background.

The processes shown in Eqs.~(\ref{eq:3tb})-(\ref{eq:4t}) provide a possible origin of this excess.
The various contributions of heavy scalars and the right-handed neutrino to the two signal regions are shown in Table~\ref{tab:ssdlbjet}
for $\tan\beta = 1$ in the upper block and $\tan\beta = 2$ in the lower block. 
We set $M_{H^\pm} = M_{H^0} = M_{A^0} = 500$ GeV  and 100\% branching fractions of the heavy scalars into $t\bar t$ or $tb$. 
We estimate the efficiency by folding in the branching fractions to contributing final states, including combinatorial factors, 
and $b$-tagging efficiencies (using 70\% for a single $b$ tag). 
To obtain the predicted number of signal events in 20 fb$^{-1}$ of data, we use a rough estimate of the  acceptance, $A= 50\%$. 
For both values of $\tan\beta$, the expected number of signal events is compatible for both signal regions
with the observed deviations above the background.
We note that the $W'$-produced events include more leptons with positive charge than with negative charge; the same feature is seen 
in the ATLAS events (Tables 11 and 12 of \cite{Aad:2015gdg}).

 \begin{table}[t]
\centering \renewcommand{\arraystretch}{1.6}
\begin{tabular}{c|c|c|c|c|r}
Signal & channel  & efficiency &  \multicolumn{2}{|c|}{signal events} & \hspace*{-2mm} obs. \hspace*{-1mm}(background) \hspace*{-3mm} \\ \hline\hline
\multirow{3}{*}{ $bb \ell^\pm\ell^\pm $}
& $H^\pm H^0,H^\pm A^0 \, \to \, 3t+b$ & $2.5\times 10^{-4}$ & 1.0-1.8 & \multirow{3}{*}{4.2-6.5} & \multirow{3}{*}{12~(4.3$\pm1.1\pm1.1)\!\!$}\\ \cline{2-4}
& $\tau N^\tau, eN^\tau \to  (\tau/e)(\tau/e) tb$ & $5.2\times 10^{-4}$ & 2.1-3.7 & \\ \cline{2-4}
& $t\bar t A^0, t\bar t H^0 \, \to \, 4t$ & $1.6\times 10^{-2}$ &1.1 & \\ \hline
\multirow{2}{*}{$\!\!\!\ge \!3b\,\ell^\pm\ell^\pm \! $} & $H^\pm H^0,H^\pm A^0 \, \to \, 3t+b$ & $6.2\times 10^{-4}$ & 2.5-4.4 & \multirow{2}{*}{5.1-7.0} & \multirow{2}{*}{6~(1.1$\pm0.9\pm0.4$)}\\ \cline{2-4}
& $t\bar  t A^0, t\bar t  H^0 \, \to \, 4t$ & $4.1\times 10^{-2}$ & 2.6 &  \\ 
\hline  \\ [-6.mm]  \hline\hline
\multirow{5}{*}{$bb \ell^\pm\ell^\pm$} & $H^\pm H^0,H^\pm A^0 \, \to \, 3t+b$ & $1.6\times 10^{-4}$ & 0.7-1.1 & \multirow{5}{*}{4.2-7.2} & \multirow{5}{*}{12~(4.3$\pm1.1\pm1.1)\!\!$}\\ \cline{2-4}
& $WH^0,~WA^0 \, \to \,  Wt\bar t$ & $2.2\times 10^{-4}$ & 0.9-1.6 & \\ \cline{2-4}
& $H^\pm h,H^\pm Z \, \to \, \ell tb + X$ & $0.7\times 10^{-4}$ & 0.3-0.5 & \\ \cline{2-4}
& $\tau N^\tau, eN^\tau \to  (\tau/e)(\tau/e)tb$ & $5.2	\times 10^{-4}$ & 2.1-3.7 & \\ \cline{2-4}
& $t\bar t A^0, t\bar t H^0 \, \to \, 4t$ & $1.6\times 10^{-2}$ & 0.3 & \\ \hline
\multirow{2}{*}{$\!\!\!\ge \!3b \, \ell^\pm\ell^\pm \! $} & $H^\pm H^0,H^\pm A^0 \, \to \, 3t+b$ & $4.0\times 10^{-4}$ & 1.6-2.8 & \multirow{2}{*}{2.3-3.5} & \multirow{2}{*}{6~(1.1$\pm0.9\pm0.4$)}\\ \cline{2-4}
& $t\bar  t A^0, t\bar t H^0 \, \to \, 4t$ & $4.1\times 10^{-2}$ & 0.7 & \\
\hline

\end{tabular}
\caption{Contributions from $W^\prime$ cascade decays and $t\bar t A^0/H^0$ production 
to the same-sign leptons plus $b$-jets signals at the 8 TeV LHC. 
The last column gives the observed and expected number of events in the ATLAS search \cite{Aad:2015gdg}. 
For the upper and lower blocks of the table $\tan\beta = 1$ and 2, respectively.
The range of predicted signal events corresponds to the 200--350 fb range for the $W'$ production rate.
The parameters are fixed as follows: heavy scalar masses are 500 GeV, the $N^\tau$ mass is 1 TeV, $s_{\theta_e} = 0.5$, and $M_{W'} = 1.9$ TeV. }
\medskip\medskip
\label{tab:ssdlbjet}
\end{table}

If the heavy scalar masses are decreased to $M_{H^\pm} \approx 400$ GeV, the number of signal events in Table 1 increases by a factor of about 3, 
so that the number of predicted events becomes too large (at $\tan\beta =2$ it may still be acceptable, given the uncertainties in the event selection efficiencies). For $M_{H^\pm} \gtrsim 700$ GeV the number of predicted events becomes too small to account for the ATLAS excess.
Thus, the preferred mass range for the heavy Higgs bosons is 400--700 GeV.

A CMS search \cite{Khachatryan:2014doa} in a similar final state with same-sign leptons and $b$ jets has yielded a smaller excess.
The sum over the number of events with two or more $b$ jets, large $H_T$ and high lepton $p_T$ 
from Table 3 of \cite{Khachatryan:2014doa} gives 11 observed events for a background of roughly $6\pm 2$ events.
The compatibility of the ATLAS and CMS results in these channels needs further scrutiny. We note though 
that the CMS event selection includes a veto of a third lepton, while the  events reported by ATLAS include events with three leptons.\footnote{In addition, the CMS jets are required to have $p_T > 40$ GeV, while for the ATLAS search a  jet $p_T > 25$ GeV cut is imposed. 
Note that the energy released in the $W'$  decay is shared between 10 or more objects, so that the stronger jet $p_T$ cut imposed by CMS may
reduce the number of observed  events.}

Searches for final states with 3 leptons may also test the presence of the heavy Higgs particles, as follows from the processes in
 Eqs.~(\ref{eq:3tb})-(\ref{eq:4t}). The CMS search \cite{Chatrchyan:2014aea} for events with 3 leptons and one or more $b$ jets 
 has yielded a deficit of events compared to the SM prediction. In particular, the large $H_T$ category with no hadronic $\tau$ decays,
 $\met > 100$ GeV and no $e^+e^-$ or $\mu^+\mu^-$ pairs  
 includes a single observed  event for an expected background of $5.5\pm 2.2$ events
 (Table 3 of  \cite{Chatrchyan:2014aea}).
Let us estimate the contribution to this event category from the $3t+b$ final state produced in 
 the $W'$ cascade decays of Eq.~(\ref{eq:3tb}). Of the  1--1.8 signal events from the first row of  Table 1,
 only a fraction of $B(W\to \ell+\met)/4 \approx 6.4\%$ would pass the criterion of 3 leptons and no $e^+e^-$ or $\mu^+\mu^-$ pairs.
 Similar suppressions apply to the number of $3 \ell$ events contributed by the processes listed in the other rows  of  Table 1.
 Thus, searches in 3-lepton final states do not appear to be able for now to differentiate between the SM and our theory with a $W'$ and 
 heavy Higgs bosons.
 
The event selection employed in the four-top ATLAS search~\cite{Aad:2015gdg} has a few additional categories, including those with lower $H_T$, one $b$-jet, lower $\met$, which have not lead to significant excesses as the categories ``SR4t3'' and ``SR4t4'' discussed above. The heavy Higgs contributions from $W^\prime$ decays to these other categories are much smaller. For instance, the heavy $W^\prime$ renders the $H_T$ large ($>700~\gev$), with harder $\met$ ($>100~\gev$). Due to the use of ``loose'' $b$-tagging in the search and the presence of four $b$ quarks in the heavy Higgs signal, the number of events in the one $b$-jet categories is a factor of 3--5 smaller than in the two $b$-jet (``SR4t3'') and three-or-more $b$-jet (``SR4t4'') categories. 

\bigskip
 
\section{Conclusions and outlook} 

A $W'$ boson of mass near 1.9 TeV with properties detailed in Ref.~\cite{Dobrescu:2015qna} provides  a compelling explanation
for the excess events reported by the ATLAS and CMS Collaborations \cite{Aad:2015owa}-\cite{Khachatryan:2014gha}
in the $WZ$, $Wh^0$, $e^+e^-jj$, and $jj$ channels.
In this article we have shown that the gauge structure associated with that $W'$ boson predicts 
the existence of heavy Higgs bosons of masses between the weak scale and a few TeV. 
We have derived the branching fractions for all $W'$ decays, including six channels with heavy Higgs bosons (see Figures 1 and 2).

The main decay modes for the heavy Higgs bosons are $A^0, H^0 \to t\bar t$ and $H^+ \to t \bar b$. If their masses are below 
$M_{W'}/2$, then the cascade decay $W' \to H^\pm A^0/H^0 \to 3t+b \to 3W+ 4b$ has a branching fraction of up to 3\% 
and provides a promising way for discovering all these particles. 
An excess of events, with a statistical significance of about $3\sigma$, has been reported by the ATLAS 
Collaboration \cite{Aad:2015gdg} in the final state with two leptons of same charge and two or more $b$ jets. We have shown that 
this can be explained by the cascade decays of $W'$ if the heavy Higgs bosons have masses in the 400--700 GeV range.
 
The $SU(2)_L\times SU(2)_R \times U(1)_{B-L}$
gauge theory presented here depends on only a few parameters, whose ranges are already determined by accounting for the 
deviations from the SM mentioned above. The various phenomena predicted by this gauge theory can thus be 
confirmed or ruled out in the near future. In Run 2 of the LHC, the $W'$ production cross section is large, in the 1--2 pb range 
at $\sqrt{s} = 13$ TeV.  Besides resonant production of $WZ$, $Wh^0$, $jj$ and $tb$,  there are several 
$W'$ discovery modes:  $W' \to \tau N^\tau \to \tau\tau jj ,  \tau\tau t b , e\tau jj, e\tau t b$ 
and $W' \to e  N^\tau \to  eejj , eetb$ would test the existence 
of the heavy right-handed neutrino, while $W' \to H^\pm A^0/H^0 \to 3t+b$,  $W' \to W A^0/H^0 \to Wt\bar t$,  $W' \to H^\pm h^0 \to tb h^0$ 
and others would test the existence of the heavy Higgs bosons. 

Another promising search channel for the heavy neutral Higgs bosons, independent of the $W'$, follows from
production in association with a $t \bar t$ pair, which has a cross section of 
the order of 10 fb at $\sqrt{s} = 13$ TeV. 
With more data, the $Z'$ boson analyzed in \cite{Dobrescu:2015qna}
will also be accessible in a variety of channels. 

\bigskip\bigskip\bigskip

{\it \bf Acknowledgments:} We would like to thank Patrick Fox, Robert Harris, Matthew Low and Andrea Tesi for helpful conversations. We would like to thank Tongyan Lin and the referee for drawing our attention to some errors in an earlier version of Section 3.
ZL was supported by the Fermilab Graduate Student Research Program in Theoretical Physics.

%

\vfil

\begin{thebibliography}{99} \frenchspacing

\bibitem{Aad:2015owa} 
  G.~Aad {\it et al.}  [ATLAS Collaboration],
  ``Search for high-mass diboson resonances with boson-tagged jets in $pp$ collisions at $\sqrt{s}$ = 8 TeV," 
  arXiv:1506.00962 [hep-ex].\\
  In a similar search, CMS reported a smaller excess:
  V.~Khachatryan {\it et al.} [CMS Collaboration],
  ``Search for massive resonances in dijet systems containing jets tagged as W or Z boson decays in pp collisions at $ \sqrt{s} $ = 8 TeV,''
  JHEP {\bf 1408}, 173 (2014)
  [arXiv:1405.1994 [hep-ex]].
  
\bibitem{Khachatryan:2014dka} 
  V.~Khachatryan {\it et al.}  [CMS Collaboration],
  ``Search for heavy neutrinos and $\mathrm {W}$ bosons with right-handed couplings in $pp$ collisions at $\sqrt{s} = 8\,\text {TeV} $,''
  Eur.\ Phys.\ J.\ C {\bf 74}, no. 11, 3149 (2014)
  [arXiv:1407.3683 [hep-ex]].

\bibitem{CMS:2015gla} 
  CMS Collaboration,
  ``Search for massive $WH$ resonances decaying to $\ell \nu b \bar{b}$ final state in the boosted regime at $\sqrt{s}=8$ TeV,''
  note PAS-EXO-14-010, March 2015.

\bibitem{Khachatryan:2015sja} 
  V.~Khachatryan {\it et al.}  [CMS Collaboration],
  ``Search for resonances and quantum black holes using dijet mass spectra in $pp$ collisions at $\sqrt{s} =$ 8 TeV,''
  Phys.\ Rev.\ D {\bf 91}, no. 5, 052009 (2015)
  [arXiv:1501.04198].\\
In a similar search, ATLAS reported a slightly smaller excess:
  G.~Aad {\it et al.} [ATLAS Collaboration],
  ``Search for new phenomena in the dijet mass distribution using $p-p$ collision data at $\sqrt{s}=8$ TeV,''
  Phys.\ Rev.\ D {\bf 91}, no. 5, 052007 (2015)
  [arXiv:1407.1376].

\bibitem{Khachatryan:2014gha} 
  V.~Khachatryan {\it et al.}  [CMS Collaboration],
  ``Search for massive resonances decaying into pairs of boosted bosons in semi-leptonic final states at $\sqrt{s} =$ 8 TeV,''
  JHEP {\bf 1408}, 174 (2014)
  [arXiv:1405.3447].\\
In a similar search, ATLAS reported a smaller excess:
  G.~Aad {\it et al.} [ATLAS Collaboration],
  ``Search for resonant diboson production in the $\mathrm {\ell \ell }q\bar{q}$ final state in $pp$ collisions at $\sqrt{s} = 8$ TeV,''
  Eur.\ Phys.\ J.\ C {\bf 75}, no. 2, 69 (2015)
  [arXiv:1409.6190].

\bibitem{Dobrescu:2015qna} 
  B.~A.~Dobrescu and Z.~Liu,
  ``A $W'$ boson near 2 TeV: predictions for Run 2 of the LHC,''
  arXiv:1506.06736 [hep-ph].
  
\bibitem{Dobrescu:2013gza} 
  B.~A.~Dobrescu and A.~D.~Peterson,
  ``$W'$ signatures with odd Higgs particles,''
  JHEP {\bf 1408}, 078 (2014)
  [arXiv:1312.1999].

  \bibitem{Jinaru:2013eya} 
  A.~Jinaru, C.~Alexa, I.~Caprini and A.~Tudorache,
  ``$W' \to hH^{\pm}$ decay in $G(221)$ models,''
  J.\ Phys.\ G {\bf 41}, 075001 (2014)
  [arXiv:1312.4268].
      
  
\bibitem{Mohapatra:1974hk} 
A very incomplete list of pointers to the vast literature on left-right symmetric models 
includes:
  R.~N.~Mohapatra and J.~C.~Pati,
  ``Left-right gauge symmetry and an isoconjugate model of CP violation,''
  Phys.\ Rev.\ D {\bf 11}, 566 (1975). \
  G.~Senjanovic and R.~N.~Mohapatra,
  ``Exact left-right symmetry and spontaneous violation of parity,''
  Phys.\ Rev.\ D {\bf 12}, 1502 (1975).
  W.~Y.~Keung and G.~Senjanovic,
  ``Majorana neutrinos and the production of the right-handed charged gauge boson,''
  Phys.\ Rev.\ Lett.\  {\bf 50}, 1427 (1983). \ 
  K.~S.~Babu and R.~N.~Mohapatra,
  ``{CP} violation in seesaw models of quark masses,''
  Phys.\ Rev.\ Lett.\  {\bf 62}, 1079 (1989).\
  E.~Ma, ``Radiative quark and lepton masses in a left-right gauge model,''
  Phys.\ Rev.\ Lett.\  {\bf 63}, 1042 (1989). \
  N.~G.~Deshpande, J.~F.~Gunion, B.~Kayser and F.~I.~Olness,
  ``Left-right symmetric electroweak models with triplet Higgs,''
  Phys.\ Rev.\ D {\bf 44}, 837 (1991). \
  J.~Choi and R.~R.~Volkas,
  ``Exotic fermions in the left-right symmetric model,''
  Phys.\ Rev.\ D {\bf 45}, 4261 (1992).
  M.~Cvetic, P.~Langacker and B.~Kayser,
  ``Determination of $g_R / g_L$ in left-right symmetric models at hadron colliders,''
  Phys.\ Rev.\ Lett.\  {\bf 68}, 2871 (1992). \
  K.~S.~Babu, K.~Fujikawa and A.~Yamada,
  ``Constraints on left-right symmetric models from the process $b \to s \gamma$,''
  Phys.\ Lett.\ B {\bf 333}, 196 (1994)
  [hep-ph/9312315]. \
  A.~Maiezza, M.~Nemevsek, F.~Nesti and G.~Senjanovic,
  ``Left-right symmetry at LHC,''
  Phys.\ Rev.\ D {\bf 82}, 055022 (2010)
  [arXiv:1005.5160]. \
  G.~Bambhaniya, J.~Chakrabortty, J.~Gluza, M.~Kordiaczyn"ska and R.~Szafron,
  ``Left-right symmetry and the charged Higgs bosons at the LHC,''
  JHEP {\bf 1405}, 033 (2014)
  [arXiv:1311.4144].
  K.~S.~Babu and A.~Patra,
  ``Higgs boson spectra in supersymmetric left-right models,''
  arXiv:1412.8714.


\bibitem{Gluza:2015goa}
  J.~Gluza and T.~Jeliński,
  ``Heavy neutrinos and the $pp \to lljj$ CMS data,''
  Phys.\ Lett.\ B {\bf 748}, 125 (2015)
  [arXiv:1504.05568 [hep-ph]].
      
      
\bibitem{Branco:2011iw} 
For a review, see   G.~C.~Branco, P.~M.~Ferreira, L.~Lavoura, M.~N.~Rebelo, M.~Sher and J.~P.~Silva,
  ``Theory and phenomenology of two-Higgs-doublet models,''
  Phys.\ Rept.\  {\bf 516}, 1 (2012)
  [arXiv:1106.0034 [hep-ph]].

\bibitem{Gunion:2002zf} 
  J.~F.~Gunion and H.~E.~Haber,
  ``The CP conserving two Higgs doublet model: the approach to the decoupling limit,''
  Phys.\ Rev.\ D {\bf 67}, 075019 (2003)
  [hep-ph/0207010].

\bibitem{Langacker:1989xa} 
  P.~Langacker and S.~U.~Sankar,
  ``Bounds on the mass of $W_R$ and the $W_L-W_R$ mixing angle $\xi$ in general $SU(2)_L \times SU(2)_R \times U(1)$ models,''
  Phys.\ Rev.\ D {\bf 40}, 1569 (1989).
  K.~Hsieh, K.~Schmitz, J.~H.~Yu and C.-P.~Yuan,
  ``Global analysis of general $SU(2) \times SU(2) \times U(1)$ models with precision data,''
  Phys.\ Rev.\ D {\bf 82}, 035011 (2010)
  [arXiv:1003.3482 [hep-ph]].

\bibitem{Brehmer:2015cia} 
  J.~Brehmer, J.~Hewett, J.~Kopp, T.~Rizzo and J.~Tattersall,
  ``Symmetry restored in dibosons at the LHC?,''
  arXiv:1507.00013 [hep-ph].
  
\bibitem{Gao:2015irw} 
  Y.~Gao, T.~Ghosh, K.~Sinha and J.~H.~Yu,
  ``G221 interpretations of the diboson and Wh excesses,''
  arXiv:1506.07511.
  Q.~H.~Cao, B.~Yan and D.~M.~Zhang,
  ``Simple non-abelian extensions and diboson excesses at the LHC,''
  arXiv:1507.00268.
  
  
\bibitem{Han:2012vk} 
  T.~Han, I.~Lewis, R.~Ruiz and Z.~g.~Si,
  ``Lepton number violation and $W^\prime$ chiral couplings at the LHC,''
  Phys.\ Rev.\ D {\bf 87}, no. 3, 035011 (2013)
  [Erratum: Phys.\ Rev.\ D {\bf 87}, no. 3, 039906 (2013)]
  [arXiv:1211.6447].

\bibitem{Blanke:2011ry} 
  M.~Blanke, A.~J.~Buras, K.~Gemmler and T.~Heidsieck,
  ``$\Delta F = 2$ observables and $B \to X_q \gamma$ decays in the Left-Right Model: Higgs particles striking back,''
  JHEP {\bf 1203}, 024 (2012)
  [arXiv:1111.5014 [hep-ph]].


    
\bibitem{Aguilar-Saavedra:2014ola} 
  J.~A.~Aguilar-Saavedra and F.~R.~Joaquim,
  ``Closer look at the possible CMS signal of a new gauge boson,''
  Phys.\ Rev.\ D {\bf 90}, no. 11, 115010 (2014)
  [arXiv:1408.2456].
  
\bibitem{Craig:2015jba} 
  N.~Craig, F.~D'Eramo, P.~Draper, S.~Thomas and H.~Zhang,
  ``The hunt for the rest of the Higgs bosons,''
  JHEP {\bf 1506}, 137 (2015)
  [arXiv:1504.04630].
  J.~Hajer, Y.~Y.~Li, T.~Liu and J.~F.~H.~Shiu,
  ``Heavy Higgs bosons at 14 TeV and 100 TeV,''
  arXiv:1504.07617.
  
  
\bibitem{Alwall:2014hca} 
  J.~Alwall, R.~Frederix, S.~Frixione, V.~Hirschi, F.~Maltoni, O.~Mattelaer, H.-S.~Shao, T.~Stelzer, P.~Torrielli, M.~Zaro, 
  ``The automated computation of tree-level and NLO differential cross sections, and their matching to parton shower simulations,''
  JHEP {\bf 1407}, 079 (2014)
  [arXiv:1405.0301 [hep-ph]].
  
\bibitem{Martin:2009iq} 
  A.~D.~Martin, W.~J.~Stirling, R.~S.~Thorne and G.~Watt,
  ``Parton distributions for the LHC,''
  Eur.\ Phys.\ J.\ C {\bf 63}, 189 (2009)
  [arXiv:0901.0002 [hep-ph]].
  
\bibitem{Aad:2015gdg} 
  G.~Aad {\it et al.}  [ATLAS Collaboration],
  ``Analysis of events with $b$-jets and a pair of leptons of the same charge in $pp$ collisions at $\sqrt{s}=8$ TeV,''
  arXiv:1504.04605.
  
\bibitem{Khachatryan:2014doa} 
  V.~Khachatryan {\it et al.}  [CMS Collaboration],
  ``Search for top-squark pairs decaying into Higgs or Z bosons in $pp$ collisions at $\sqrt{s}$=8 TeV,''
  Phys.\ Lett.\ B {\bf 736}, 371 (2014)
  [arXiv:1405.3886 [hep-ex]].
  
\bibitem{Chatrchyan:2014aea} 
  S.~Chatrchyan {\it et al.}  [CMS Collaboration],
  ``Search for anomalous production of events with three or more leptons in $pp$ collisions at $\sqrt(s) =$ 8 TeV,''
  Phys.\ Rev.\ D {\bf 90}, 032006 (2014)
  [arXiv:1404.5801 [hep-ex]].
  
\end{thebibliography}
\end{document}